\documentclass[12pt]{article}
\usepackage{graphicx}
\usepackage{booktabs}
 \usepackage{lscape}
 \usepackage{verbatim}
 \usepackage{multirow}
 \usepackage{geometry}
 \usepackage{amssymb}
 \usepackage{units}
 \usepackage{setspace}
 \usepackage{dsfont}
  \usepackage[numbers]{natbib}
   \usepackage{bm}
   \usepackage{color}
   \usepackage{amsmath}
   \usepackage{comment}
   \usepackage{graphicx,wrapfig,lipsum,multirow,subfigure}
   \usepackage{color, colortbl}
   \usepackage[nolists,tablesonly]{endfloat}

\definecolor{Gray}{gray}{0.75}
\definecolor{LightGray}{gray}{0.9}

\newcommand{\tj}[1]{\textcolor{red}{#1}}
\title{Backfilling Cohorts in Phase I Dose-Escalation Studies}

\author{Helen Barnett$^1$, Oliver Boix$^2$, Dimitris Kontos$^3$, Thomas Jaki$^{1,4}$\\$^1$ MRC Biostatistics Unit, University of Cambridge\\
$^2$ Bayer AG\\
$^3$ ClinBAY\\
$^4$ Department of Mathematics and Statistics, Lancaster University}
\begin{document}
\maketitle
\begin{abstract}
The use of `backfilling', assigning additional patients to doses deemed safe, in phase I dose-escalation studies has been used in practice to collect additional information on the safety profile, pharmacokinetics and activity of a drug. These additional patients help ensure that the Maximum Tolerated Dose (MTD) is reliably estimated and give additional information in order to determine the recommended phase II dose (RP2D). In this paper, we study the effect of employing backfilling in a phase I trial on the estimation of the MTD and the duration of the study. We consider the situation where only one cycle of follow-up is used for escalation as well as the case where there may be delayed onset toxicities. We find that, over a range of scenarios, there is an increase in the proportion of correct selections and a notable reduction in the trial duration at the cost of more patients required in the study.
\end{abstract}

\section{Introduction}
In Phase I dose-finding studies, the main objective is often to find the Maximum Tolerated Dose (MTD) or the recommended Phase II dose (RP2D), the dose recommended for further testing in Phase II. The MTD is defined as highest dose that has an acceptable level of toxicity \citep{Storer2001}, most often corresponding to a certain probability of occurrence of a Dose Limiting Toxicity event (DLT). In oncology, a DLT is frequently defined as a grade 3 or higher toxicity by the grading scale of the National Cancer Institute \citep{NCI}. In the following, we use the terms DLT and toxicity interchangeably.\\

In a Phase I dose-escalation study, a set of doses is investigated, patients are recruited in cohorts and an escalation procedure is used to carefully escalate from low doses that are expected to be very safe to dose levels that have an acceptable level of toxicity and at the same time induce some desirable activity in a patient. The escalation procedure can be rule-based \cite[e.g.][]{Storer2001}, model-based \cite[e.g.][]{wheeler2019} or model-assisted \cite[e.g.][]{Liu2015} and cohorts are most commonly small \cite{zhou2003}, with a size of 3 often used in such trials.\\

However, with such small sample sizes, comes greater uncertainty in the estimate of interest. It is, for example, desirable to establish the MTD quickly and accurately but with no more patients than necessary. As with any clinical trial, a balance must be taken between the accuracy, the duration of the study and the trial size. Typically, the larger the trial, the higher the accuracy, but also the longer the duration of the study.\\

An approach that has gained popularity in the recent years (e.g. \cite{Hamilton2019,Pauff2021}), is the use of `backfilling' of cohorts on lower doses \cite{dehbi2021}. The principle is that, once a dose is deemed safe enough to escalate to a higher dose-level, additional patients may be allocated to lower doses to increase the understanding of the safety, tolerability and activity of these doses. The decision to backfill a dose may be taken solely on the criteria that the dose is deemed `safe', or it may require the additional condition that an activity signal must be seen. \\

Whilst it is clear that backfilling will result in a, potentially large, increase in patient numbers, additional insight around the safety, tolerability and potential activity of the treatment is gained. Moreover, there is the potential that the trial duration could be substantially reduced due to the improved understanding of the dose-toxicity relationship. The exact nature of this relationship between trial duration and sample size is, however, unclear - in particular in the setting where late-onset toxicities are of concern.\\

In this work we investigate the impact of backfilling on the probability of correctly selecting the MTD and on the duration in phase I dose-escalation trials. The rest of the paper is organized as follows. In Section~\ref{sec:Mot_ex}, we give an example of a dose-escalation trial that utilized backfilling. In Section~\ref{sec:Methods}, we first introduce an example trial simulation with and without backfilling, then describe the dose-finding algorithms used in this work. In Section~\ref{sec:Results}, we present the results from simulation studies to demonstrate the impact of backfilling on operating characteristics. Finally, we conclude with a discussion in Section~\ref{sec:Discussion}.\\

\section{Motivating Trial Example} \label{sec:Mot_ex}

The first in-human Phase I study reported in \cite{tao2019} investigated the safety and activity of the activin A inhibitor, STM 434, in advanced solid tumors. The study initally considered five doses levels, 0.25, 0.5, 1, 2, and 4 mg/kg administered every 4 weeks. The treatment scheduled subsequently changed to bi-weekly after the half-life was estimated to be lower than anticipated and an additional dose of 8 mg/kg was added later due to lower than anticipated
predicted exposures. A 3+3 design \cite{Storer2001} was used to guide dose-escalation. Dose-limiting toxicities (DLT) were deﬁned as any grade $\geq$3 nonhematologic toxicity, any grade $\geq$4 hematologic toxicity lasting 7 days, febrile neutropenia, or grade 3 thrombocytopenia with active
bleeding and the DLT assessement period was 28 days. A minimum of 3 evaluable patients were required prior to dose-escalation and "Backfill" slots were permitted at doses that had been declared safe.\\

A total of 32 patients participated in the trial of which three experienced a DLT and of 28 patients that were evaluable according to RECIST \cite{RECIST}, 16 achieved stable disease while the remaining had progressive disease  (Table \ref{tab:ex}). Since two of the DLTs occurred in the highest dose, the MTD was declared to be 4 mg/kg every 2 weeks but no dose-expansion was undertaken following the safety review committees recommendation on the basis of the overall safety profile observed.

\begin{table}
 \begin{tabular}[ht]{ccccccccccc}
 \rowcolor{Gray}Dose (mg/kg) & 0.25  & 0.5 & 0.5  & 0.75 & 1 & 2 & 4 & 8 \\
 \rowcolor{Gray}& 4 weekly & 4 weekly & 2 weekly & 2 weekly& 2 weekly & 2 weekly & 2 weekly & 2 weekly\\\hline
 DLT & 0 & 1 & 0 & 0 & 0 & 0 & 0&  2\\
 SD & 1 & 3 & 3 & 1 & 2 & 3 & 2 & 1\\
 \rowcolor{LightGray}Total subjects & 4 & 6& 4 &3 & 4& 4 & 4& 3\\
\end{tabular}
\caption{Number of subjects and DLT per dose-level.\label{tab:ex}}
\end{table}

From the results of the study we can see that backfilling slots were indeed used as several of the cohorts have more than three (or six) patients required for the 3+3 design.\\ 

Note, that the 3+3 design does, however, not utilize these additional data when recommending subsequent doses. In the remainder of this work we will therefore explore the impact of backfilling on the operating characteristics of a trial that utilizes a model-based escalation approach. This exploration is motiviated by a recent study of Thorium-227 in combination with an antibody (NCT03507452). As this study utilized doses of Thorium-227 of 1.5 MBq in steps of 1.0 or 1.5 MBq, with antibody doses of 10 mg, we will consider the same dose-levels in our subsequent evaluations. Moreover, due to the radioactive nature of Thorium-227, late-onset toxicities $-$  toxicities that occur after the first treatment cycle $-$ are of potential concern. We therefore also explore the impact of backfilling under a traditional CRM-type model as well as a time-to-event model.\\

\section{Methodology} \label{sec:Methods}
In this section we outline the methods used in this work. In Section~\ref{sec:single} we illustrate and discuss the concept of backfilling in more detail. We then describe the statistical models used for dose-escalation in Section~\ref{sec:CRM}, and finally outline the rules used in the design implementation in Section~\ref{sec:stopping}.\\

\subsection{Single Simulation Examples} \label{sec:single}
To illustrate the concept of backfilling, and the difference between trials that use backfilling and those that do not, we present an example of a single trial simulation. Following the motivating study (NCT03507452) the following six doses are investigated: 1.5MBq , 2.5MBq , 3.5MBq , 4.5MBq , 6.0MBq , 7.0MBq. Cohorts of size three enter the trial, starting at the lowest dose. If no backfilling is implemented, then cohorts are assigned to doses according to the escalation procedure, until some stopping rule is triggered. If full backfilling is implemented then, when a dose is considered safe enough that the escalation continues above it, two additional backfilling cohorts are recruited to that dose. Of course one may also choose to backfill more cautiously, but in this illustration we use two cohorts as standard. Full backfilling therefore means that we start backfilling from the lowest dose of 1.5MBq. Patients are followed up for three cycles of treatment, with new cohorts assigned after each cycle, with a cycle lasting 6 weeks. Since the follow-up period is longer than the times between assignments, the Time-to-Event Continual Reassessment (TITE-CRM) is implemented for the dose-escalation procedure as described in Section~\ref{sec:CRM}.\\

\begin{figure}
      \includegraphics[width=1\linewidth]{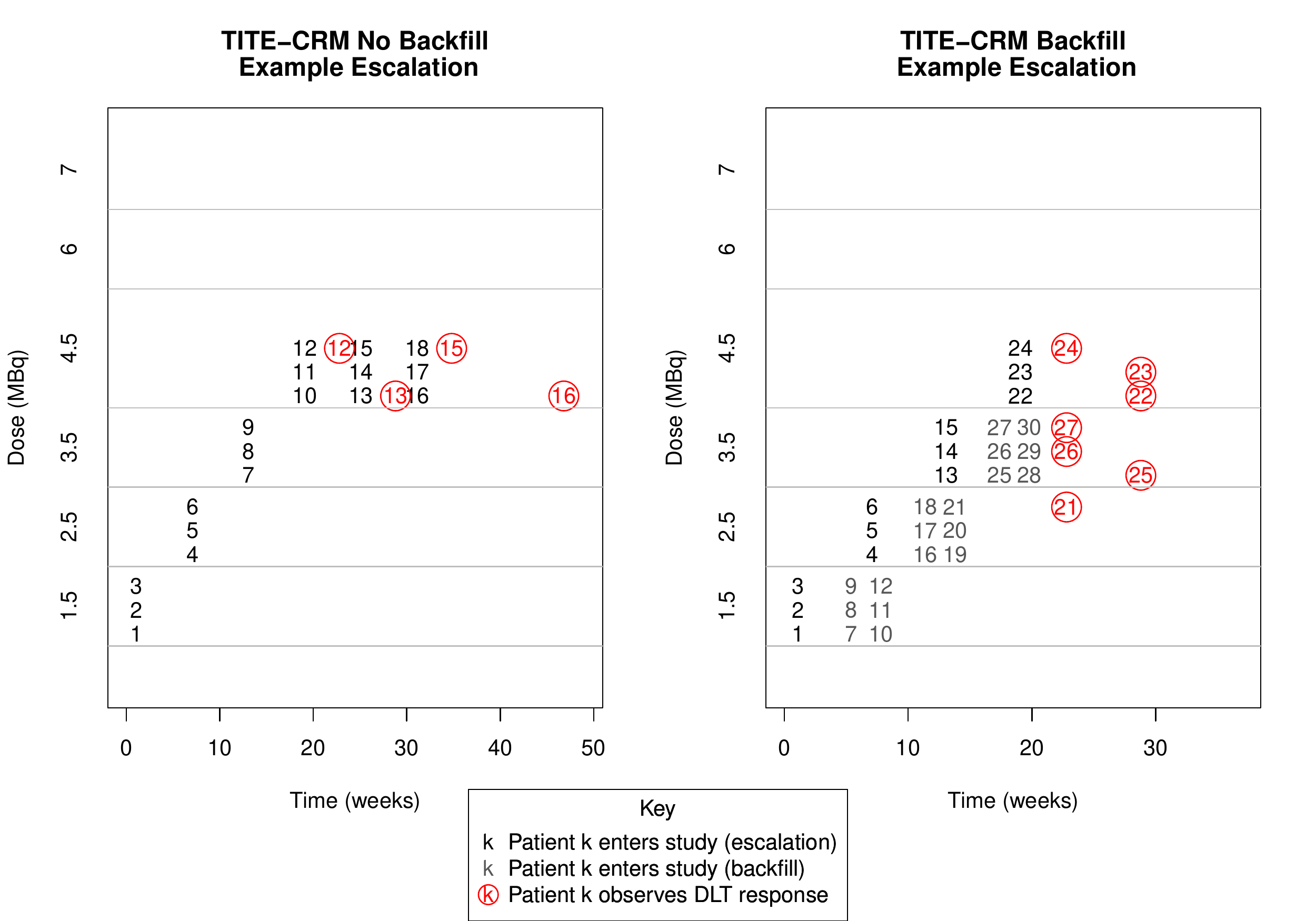}
     \caption{A single trial simulation using the TITE-CRM, with and without backfilling. Note that the backfilled cohorts in grey all enter the trial at the same time as the escalation cohort above in black, the horizontal offset is for ease of reading.}
     \label{fig:single_run}
  \end{figure}

Figure~\ref{fig:single_run} illustrates a trial is conducted without backfilling (left panel) and with full backfilling (right panel). Every other aspect of the trials was comparable (see Section~\ref{sec:stopping} for further details).  In this case, the trial without backfilling determined the MTD to be the 4.5MBq dose while the trial using backfilling recommended 3.5MBq. Figure~\ref{fig:single_run} shows that the escalation in both settings is the same - they both reach the 4.5MBq dose, but in the trial without backfilling two further cohorts are allocated to this dose before being recommended as the MTD. When backfilled cohorts were used, more information was available at the lower doses allowing the precision stopping rule to be reached quickly and hence the lower dose is recommended. In addition to the difference in recommended dose one can also observe that the duration of the study is notably shorter when backfilling is used.\\

\subsection{BLRM and TITE-BLRM} \label{sec:CRM}
In order to investigate the impact of backfilling on the estimation of the MTD and the duration of the trial, we consider two settings. The first setting assumes that patients are followed up for one cycle of treatment only (i.e. DLT period 1 cycle), and the next cohort of patients is assigned once the previous cohort's follow-up period has been fully observed. The second setting assumes that there may be late onset toxicities. Patients are therefore followed up for three cycles of treatment. A new cohort of patients is admitted every cycle, so that only partial information is available for the previous two cohorts, as their full follow-up period has not yet been observed. In the first setting we use a Baysian Logistic regression model (BLRM)  \cite{Neuenschwander2008}, and in the second setting a Time-To-Event version of the BLRM (TITE-CRM) \cite{YingKuenCheung2000}.\\

In each setting, a set of $J$ doses labelled $d_j$ for $j=1, \ldots , J$ are investigated, with patients labelled $i=1, \ldots , n$. \\

The BLRM is conducted in the following way. Starting at the lowest dose $d_1$, cohorts of patients enter the trial. After each cohort has been fully observed, the dose assignment of the next cohort is decided. A two-parameter logistic model is used to describe the dose-response relationship:

\[
F(d,\bm{\beta})= \frac{\exp(\beta_0 + \beta_1 d)}{1+\exp(\beta_0 + \beta_1 d)}, \label{eq:crm}
\]

where $F(d,\bm{\beta})$ is the probability of DLT at dose $d$, and $\bm{\beta} = (\beta_0 , \beta_1)$ is parameter vector with prior:

\[
\begin{pmatrix}
\beta_0 \\
\log (\beta_1)
\end{pmatrix} \sim N_2 \left( \begin{pmatrix}
c_1\\
c_2
\end{pmatrix} , \begin{pmatrix}
v_1 & 0 \\
0 & v_2 
\end{pmatrix} \right).
\]
The posterior distribution of $\bm{\beta}$ is updated after each cycle using the likelihood
\[
{\mathcal{L}(\bm{\beta}) = \prod_{i=1}^{n} F(d,\bm{\beta})^{y_{i}} \{ 1-F(d,\bm{\beta})\}^{1-y_{i}}},
\]
where $y_i$ is an indicator taking the value 1 if patient $i$ had observed a DLT response and 0 otherwise. The updated posterior for $\bm{\beta}$ is then used to estimate the probability of DLT at each dose. The dose assignment of the next cohort is then dose $d_j$ that minimizes $|F(d_j, \hat{\bm{\beta}}) - \tau_1 |$, where $\tau_1$ is the target DLT rate for one cycle of follow-up, subject to certain rules - see Section~\ref{sec:stopping}. The final dose recommendation is then the dose $d_j$ that minimizes$|F(d_j, \bm{\beta}) - \tau_1 |$ once a stopping rule has been triggered.\\

In the second setting, the TITE-CRM \citep{YingKuenCheung2000} is used. Here, patients are followed up for three cycles of treatment. However, if it was required to wait until the entire follow-up for the previous cohort had been completely observed to assign the dose for the next patient, the trial length would be very undesirably long. Therefore, each new cohort is assigned their dose once the previous cohort has been observed for one cycle of treatment. There is therefore only partial information available for the previous two cohorts. The TITE-CRM takes this into account by weighting the observations in the following way.\\

The dose response model $F(d,\bm{\beta})$ is identical to (\ref{eq:crm}), and is weighted to form $G(d,w,\bm{\beta})$:

\[ 
{G(d,w,\bm{\beta}) = w F(d,\bm{\beta)}},
\]
where the weights {$0 \leq w \leq 1$} are a function of time-to-event of a patient response. The posterior distribution of $\bm{\beta}$ is updated after each cycle using likelihood
\[
{\mathcal{L}(\bm{\beta}) = \prod_{i=1}^{n} G( d_{[i]}, w_{i,n}, \bm{\beta})^{y_{i,n}} \{ 1-G( d_{[i]}, w_{i,n}, \bm{\beta})\}^{1-y_{i,n}}},
\]
where $n$ is the number of patients that have been treated so far, $d_{[i]}$ is the dose assigned to patient $i$ and $y_{i,n}$ is an indicator which takes the value 1 if patient $i$ has observed a DLT after the $n$ patients treated so far have been observed for at least one cycle, and 0 otherwise.\\

The updated posterior for $\bm{\beta}$ is used to estimate the probability of DLT at each dose, as in the one cycle setting. The dose assignment of the next cohort is then the dose $d_j$ that minimizes $|F(d_j, \bm{\beta}) - \tau_3 |$, where $\tau_3$ is the target DLT rate for three cycles of follow-up, again subject to certain rules - see Section~\ref{sec:stopping}.\\

In this implementation, we use the simple specification of weights suggested by Cheung and Chappell \cite{YingKuenCheung2000}: $w_{i,n}=u_{i,n}/S$, where $u_{i,n}$ is the current number of cycles for which patient $i$ has been observed and $S$ is the total number of cycles in the follow-up period. If patient $i$ observes a DLT response, then $w_{i,n}=1$. The final dose recommendation is the dose level that minimises $|F(d_j, \hat{\beta}) - \tau_3 |$ once a stopping rule has been implemented and the follow-up for all enrolled patients has been completed.\\

Due to the added complexity of the method, the TITE-CRM uses a start-up period such that the dose assignment is escalated one level at a time until a DLT response is observed. Once a DLT response is observed, then the TITE-CRM model is used. Although in the original description of the methodology, this start-up period requires each patient to be followed up for their entire follow-up time before the next patient's dose is assigned, in our implementation, only one cycle is required for follow-up before the next is assigned, which is in line with the rest of the trial.\\

\subsection{Rules} \label{sec:stopping}
To evaluate the performance in the setting of a Phase I dose-escalation trial, we use enforcement and stopping rules that could be used in such a trial. We define $p_{s,d_j}$ as the $P(DLT)$ in cycles up to and including cycle $s$ for any given dose $d_j$.\\

\textbf{\textit{Enforcement Rules:}}
\begin{enumerate}
\item \textbf{Hard Safety}: If there is a high probability that the toxicity of an experimented dose exceeds the target toxicity, this and all higher doses are excluded from further experimentation (i.e. dose $d_j$ and all above are excluded when $P(p_{1,d_j}>\tau_1)>\psi$ for some threshold $\psi$). In this implementation we use a threshold for excessive toxicity of $\psi =0.95$, with a $Beta(1,1)$ prior for the DLT outcome. For example for $\tau_1=0.3$, if there are at least 3 DLT responses out of 3 patients, or at least 4 DLT responses out of 6 patients, or at least 5 DLT responses out of 9 patients, then all dose assignments must be lower than that dose for the rest of the study. If the lowest dose is excluded then the trial stops with no dose recommendation made.
\item \textbf{K-fold Skipping Doses}: No more than a 2-fold-rise in dose value for the next dose assignment based on the highest experimented dose so far.
\end{enumerate}

\textbf{\textit{Stopping Rules:}}
\begin{enumerate}
\item \textbf{Sufficient Information}: If a dose is recommended for the next cohort on which three cohorts have already been assigned in the escalation (excluding backfilling cohorts) the trial is stopped.
\item \textbf{Lowest Dose Deemed Unsafe}: If $P(p_{1,d_1}>30\%)>0.80$ according to the escalation model and at least one cohort of patients has been assigned to dose $d_1$, the trial is stopped.
\item \textbf{Highest Dose Deemed Very Safe}: If $P(p_{1,d_J}\leq 30\%)>0.80$ according to the model and at least one cohort of patients has been assigned to dose $d_J$, the trial is stopped.
\item \textbf{Precision}: If the MTD is estimated precisely enough, the trial is stopped. This precision is defined as $CV(MTD)<30\%$, with the coefficient of variation calculated as an adjusted median absolute deviation divided by the median. This stopping rule is only used once at least three cohorts of patients that are not part of a backfilling cohort have had at least one cycle of treatment in the escalation. The data of the backfilling cohorts are included in the estimation of the precision, however.
\item \textbf{Hard Safety}: If the lowest dose is considered unsafe according to the hard safety enforcement rule, the trial is stopped.
\item \textbf{Maximum Patients}: If the maximum number of patients ($n=n_{\mbox{max}}$) have been recruited.
\end{enumerate}

\section{Simulations} \label{sec:Results}

\subsection{Set-up}
In order to investigate the effect of backfilling on the operating characteristics of the dose-finding designs, we conduct a simulation study. The same set of six doses as used in the example in Section~\ref{sec:single} are used: 1.5MBq , 2.5MBq , 3.5MBq , 4.5MBq , 6.0MBq , 7.0MBq. Seventeen scenarios are considered, to cover a wide range of potential dose responses. Table~\ref{tab:scenarios} gives the probability of a DLT in the first six week cycle for each of the considered scenarios, with the MTD associated with the target, $\tau_1=0.3$, highlighted in boldface. Scenarios 1 -- 6 represent cases where each level in turn is the MTD, with higher and lower doses equidistant in terms of probability of DLT. Scenarios 7 and 8 are non-linear around the MTD. In scenarios 9 -- 12, no dose is exactly on target, with all doses unsafe in scenario 9. Scenarios 13 -- 17 are a set of varying scenarios, typically used to test performance of a dose-finding algorithm.\\

\begin{table}[ht]
\centering
\begin{tabular}{rrrrrrr}
  \hline
 \rowcolor{Gray}Scenario & 1.5MBq & 2.5MBq & 3.5MBq & 4.5MBq & 6.0MBq & 7.0MBq \\ 
  \hline
1 & \textbf{0.30} & 0.40 & 0.50 & 0.60 & 0.70 & 0.80 \\ 
 \rowcolor{LightGray} 2 & 0.20 & \textbf{0.30} & 0.40 & 0.50 & 0.60 & 0.70 \\ 
  3 & 0.10 & 0.20 & \textbf{0.30} & 0.40 & 0.50 & 0.60 \\ 
 \rowcolor{LightGray} 4 & 0.05 & 0.10 & 0.20 & \textbf{0.30} & 0.40 & 0.50 \\ 
  5 & 0.05 & 0.10 & 0.15 & 0.20 & \textbf{0.30} & 0.40 \\ 
 \rowcolor{LightGray} 6 & 0.02 & 0.05 & 0.10 & 0.15 & 0.20 & \textbf{0.30} \\ 
  7 & 0.15 & 0.20 & 0.25 & \textbf{0.30} & 0.45 & 0.60 \\ 
 \rowcolor{LightGray} 8 & 0.05 & 0.15 & \textbf{0.30} & 0.35 & 0.40 & 0.45 \\ 
  9 & 0.40 & 0.45 & 0.50 & 0.55 & 0.60 & 0.65 \\ 
 \rowcolor{LightGray} 10 & 0.05 & 0.15 & \textbf{0.25} & 0.35 & 0.45 & 0.55 \\ 
  11 & 0.15 & \textbf{0.20} & 0.35 & 0.40 & 0.45 & 0.50 \\ 
 \rowcolor{LightGray} 12 & 0.05 & 0.10 & 0.15 & 0.20 & \textbf{0.25} & 0.40 \\ 
  13 & 0.06 & 0.07 & 0.08 & 0.09 & 0.11 & 0.12 \\ 
 \rowcolor{LightGray} 14 & 0.10 & 0.14 & 0.21 & \textbf{0.30} & 0.46 & 0.58 \\ 
  15 & 0.16 & \textbf{0.30} & 0.50 & 0.70 & 0.89 & 0.95 \\ 
 \rowcolor{LightGray} 16 & 0.55 & 0.91 & 0.99 & 1.00 & 1.00 & 1.00 \\ 
  17 & 0.05 & 0.05 & \textbf{0.05} & 0.80 & 0.80 & 0.80 \\ 
   \hline
\end{tabular}
\caption{$p_{1,d_j}=P(DLT)$ in cycle 1 for each dose $d_j$ across 17 defined scenarios. MTD in \textbf{bold}.} \label{tab:scenarios}
\end{table}

For the setting where the follow-up period is three cycles, the conditional probability of DLT response in subsequent cycles is multiplied by a factor of 1/3. So that 
\[
p_{2,d_j}=p_{1,d_j} + (1-p_{1,d_j})\frac{p_{1,d_j}}{3},
\]
and
\[
p_{3,d_j}=p_{1,d_j} + (1-p_{1,d_j})\frac{p_{1,d_j}}{3} + (1-p_{1,d_j})(1-\frac{p_{1,d_j}}{3})\frac{p_{1,d_j}}{9}.
\] 
For example if $p_{1,d_j}=0.3$, then $p_{2,d_j}=0.37$ and $p_{3,d_j}=0.391$, hence we use $\tau_3=0.391$ as the target toxicity for three cycles of follow-up.\\

The maximum sample size is chosen to be $n_{\mbox{max}}=54$ which is relatively large for a Phase I trial, but has been chosen to allow escalation to dose $d_6$ with backfilling implemented.\\

We conduct 5,000 simulations for each scenario, and compare the performance to the non-parametric benchmark \cite{Oquigley2002}. This benchmark gives an indication of the `difficulty' of a scenario, so that we can quantify any differences between the approaches accordingly. 
Note that this uses the maximum sample size in every simulation, and is not subject to any stopping rules.\\

In these implementations, as well as comparing backfilling all doses considered safe (Fully Backfilled), and not (Not Backfilled), we also consider the setting where a dose level is only backfilled once an activity signal has been observed at that dose or any dose below (Partially Backfilled). For example, if an activity signal is first seen in the third dose level and not the first or second, then the decision to escalate to the fourth dose level would mean the additional backfilling cohorts are assigned to dose three but not doses one and two. In this partially backfilling setting, the underlying probability of observing an activity signal (at least one complete response in a cohort) is 0.00, 0.15, 0.30, 0.45, 0.60, 0.75 at each of the six doses respectively. \\

\subsection{Prior Specification}
We consider two options for the values of the hyper-parameters of the prior distribution. In the first option, the hyper-parameters for the prior used are: $c_1=\log (0.5) $, $c_2=0$, $v_1=4$ and $v_2=1$, chosen so that the prior is relatively vague, with a mean effective prior sample size of 1.3 patients per dose level. These choices are in line with those used by Neuenschwander et al. \cite{Neuenschwander2008}. Importantly, these hyper-parameters are the same for the implementation with and without backfilling, and for the follow-up of one and three cycles. The decision to choose the same prior for both one cycle and three cycles is to ensure a fair comparison between the two approaches. It is important, however, to note that in practice, different prior distributions are likely suitable for the different settings, as explored further below. \\

An alternative option considered here is to calibrate the values of these hyper-parameters over a small range of scenarios, to choose the values yielding the best performance across a wide range of settings (see for example Mozgunov et al. \cite{Mozgunov2021}). We have chosen scenarios 1, 3, 4, 6, 9 and 13 to represent a diverse set of dose-response relationships. The calibration is done separately for the four settings resulting in the combination of 1 cycle/3 cycles and backfilling/no backfilling, in order to give each approach the maximum chance of successful performance. It is important to note that the prior used for both fully and partially backfilling is the same in each case, and has been calibrated under the assumption of full backfilling. The values of these hyper-parameters resulting from the calibration procedure are provided in Table \ref{tab:prior}.\\

We have chosen to implement both approaches to the prior specification here as we wish to demonstrate a direct comparison between backfilling and not backfilling using the same prior for both, as well as demonstrating the potential of both approaches with a suitably matched prior. In practice it may be the case that the decision to backfill is made during the design stage, in which case we would use the corresponding calibrated prior. However, it may instead be the case that backfilling is introduced later in the trial (per amendment), and the same specification of prior that was intended to be used without backfilling must still be used. Hence we highlight the importance of investigating the performance for both a calibrated and non-calibrated prior.\\

\begin{table}[ht]
\centering
\begin{tabular}{ccc}
  \hline
\rowcolor{Gray} & \textbf{1 cycle} & \textbf{3 cycles}  \\ 
  \hline
\textbf{No Backfilling} &  $c_1=\log (1/2) $, $c_2=\log (1/4)$, & $c_1=\log (1/16) $, $c_2=\log (1/16)$,\\& $v_1=4$ and $v_2=2$ &  $v_1=1$ and $v_2=1$  \\ 
\rowcolor{LightGray}\textbf{Backfilling} & $c_1=\log (1/2) $, $c_2=\log (1/4)$, & $c_1=\log (1/16) $, $c_2=\log (1/16)$,\\ \rowcolor{LightGray}\textbf{(Partial \& Full)}& $v_1=2$ and $v_2=1$ & $v_1=1$ and $v_2=1$ \\

   \hline
\end{tabular}
\caption{Calibrated values for prior hyper-parameters.} \label{tab:prior}
\end{table}

\subsection{Results}
The most often considered metrics of performance in dose-finding trials are the proportion of correct selections (PCS) and the proportion of acceptable selections (PAS). A correct selection is defined as selecting the MTD as defined in Table~\ref{tab:scenarios}, or as a safety stopping rule being correctly triggered. An acceptable selection is defined as a dose whose true probability of toxicity during the first cycle is between 0.18 and 0.33. \tj{Why this range? Any reference that could be given to why?}\\

\subsubsection{Non-Calibrated Prior}

  \begin{figure}
 
      \includegraphics[width=1\linewidth]{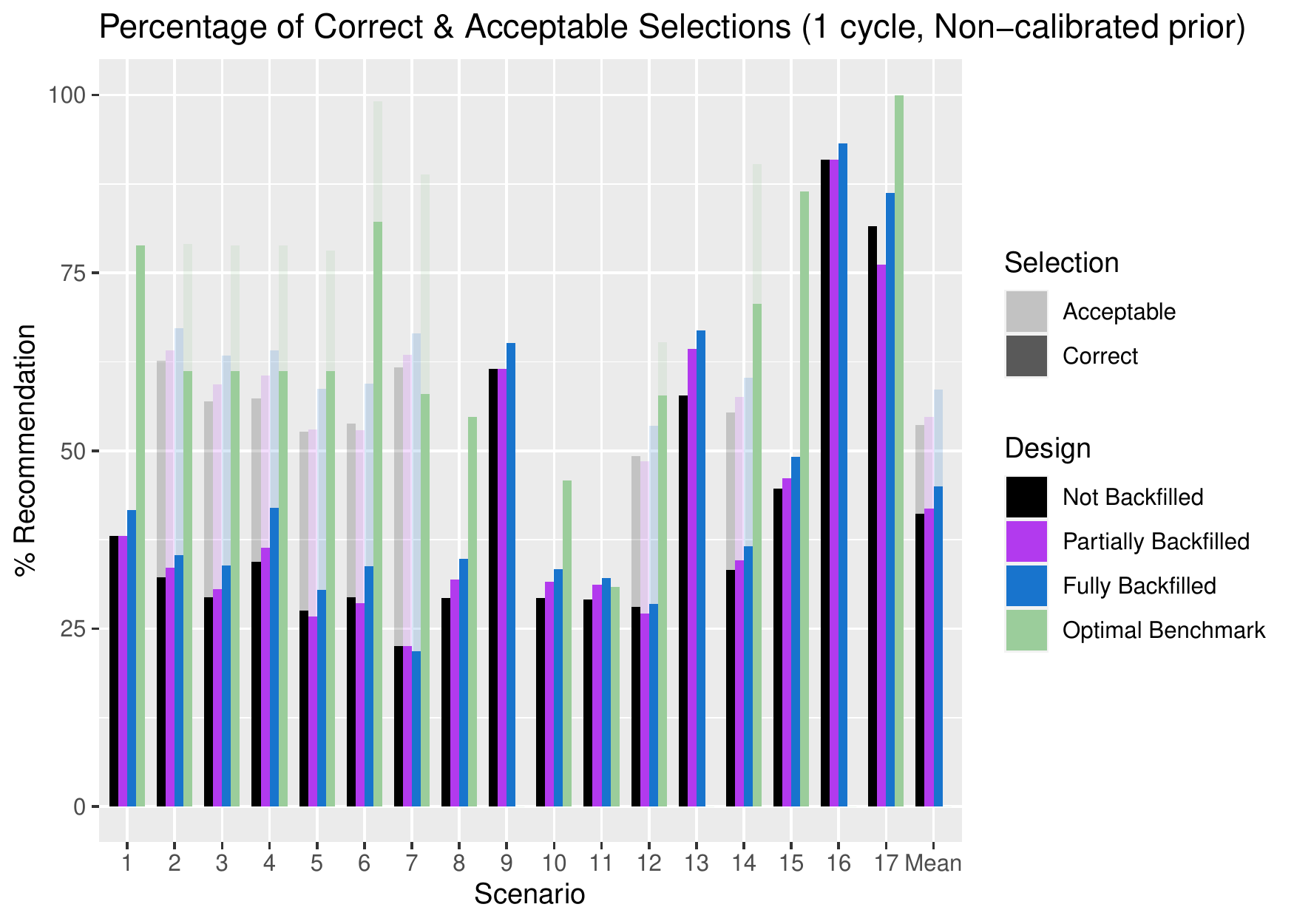}
     \caption{Proportion of correct and acceptable selections across scenarios for one cycle of follow-up}   \label{fig:PCSPAS_cyc1}
  \end{figure}

  \begin{figure}
   
      \includegraphics[width=1\linewidth]{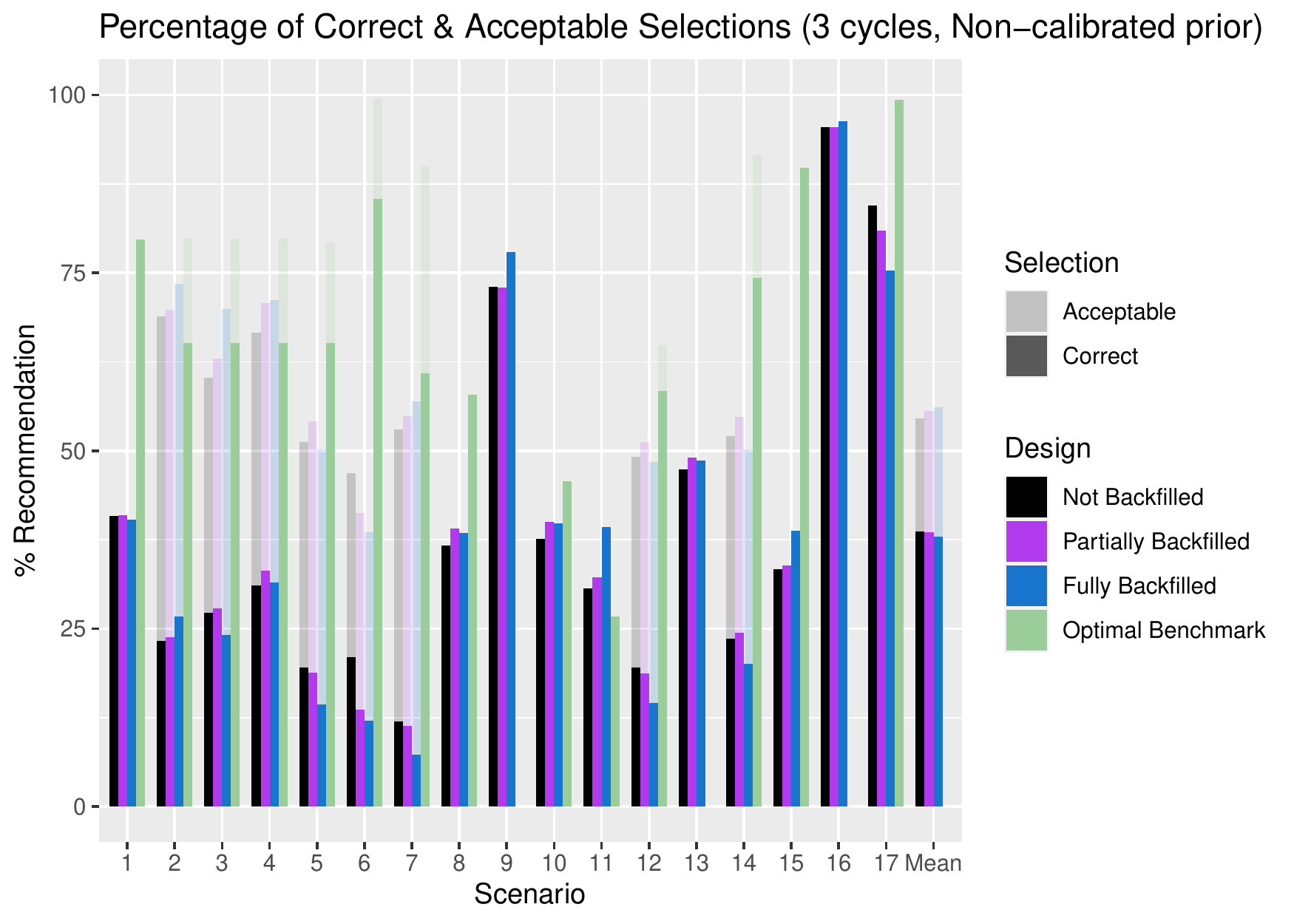}
     \caption{Proportion of correct and acceptable selections across scenarios for three cycles of follow-up} \label{fig:PCSPAS_cyc3}
  \end{figure}

Figures~\ref{fig:PCSPAS_cyc1} and \ref{fig:PCSPAS_cyc3} show the PCS and PAS in the setting with one cycle and three cycles respectively. From Figure~\ref{fig:PCSPAS_cyc1}, it can be seen that, when one cycle of follow-up is considered, employing backfilling increases the PCS and PAS in almost all scenarios, as expected. The only scenario which did not see an increase in PCS is scenario 7, where there is a marginal decrease in PCS, but a noticeable increase in PAS. This is a particularly challenging scenario, where the dose below the MTD has true probability of DLT only 5\% below target. The largest increase is seen in scenario 13, where full backfilling gives a 9\% increase in correct selections. In nearly all scenarios, partial backfilling gives a performance between no backfilling and fully backfilling as expected. Interestingly in scenario 17, the performance of partial backfilling is worse than both no and full backfilling. With such an `easy' scenario, the PCS is already very high, and the decrease is due to a shift to recommending the unsafe fourth dose. The average increase in correct selections from no backfilling to fully backfilling is 4\%, and is 5\% for acceptable selections. The approach to partially backfilling we have taken means the probability of backfilling any given level dose increases with dose. Therefore when the lower doses are more toxic, fewer doses will be backfilled and partial backfilling gives more similar results to no backfilling. Although in all cases it is noticeable that the performance is well below that of the benchmark, it is worth noting that the benchmark has the advantage of a much larger sample size on average.\\
 
\begin{figure}
  
      \includegraphics[width=1\linewidth]{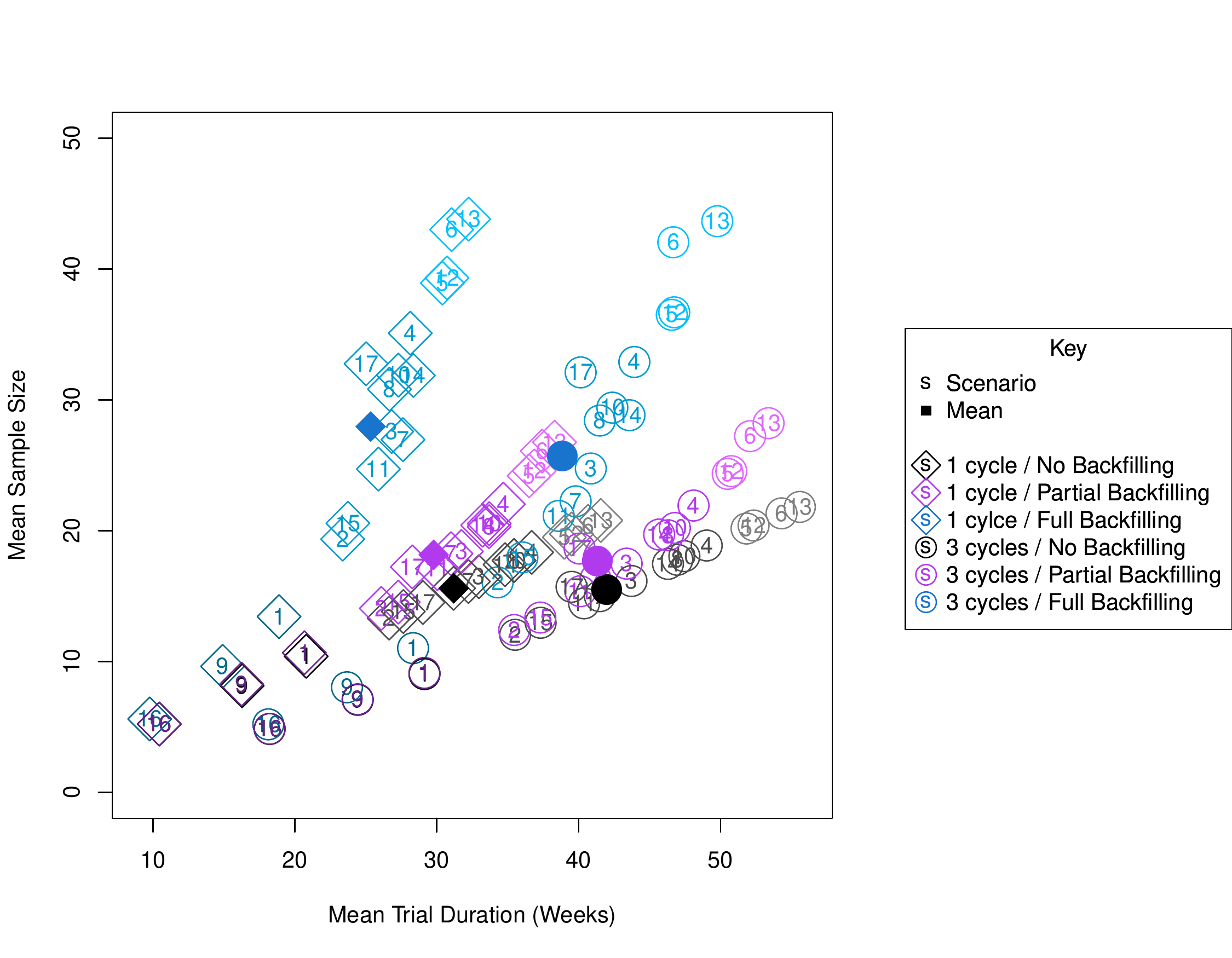}
     \caption{Mean total sample size and mean trial duration across scenarios. Darker colours indicate scenarios where a higher number of doses are unsafe.}  \label{fig:SSize_Durations}
  \end{figure}

To compare the measures of trial sizes of the two approaches, Figure~\ref{fig:SSize_Durations} shows the relationship between the mean total sample size and the mean trial duration for each scenario. The diamonds represent the setting with one cycle of follow-up, with blue indicating that full backfilling was used, purple indicating partial backfilling and black indicating no backfilling. It is clear to see that the use of full backfilling substantially increases the total sample size in scenarios where the MTD is at least the second dose level. It is notable, however, that this increase in sample size owed to the use of backfilling  decreases the average trial duration notably. On average across scenarios, full backfilling increases sample size by 12 patients on average but reduces the trial duration by 6 weeks. It appears that on average each additional patient reduces the duration by half a week. For partial backfilling, although the overall increase in sample size is smaller, as is the decrease in trial duration, the trade-off is still on average each additional patient reduces the duration by half a week. \\

\begin{figure}
   
      \includegraphics[width=1\linewidth]{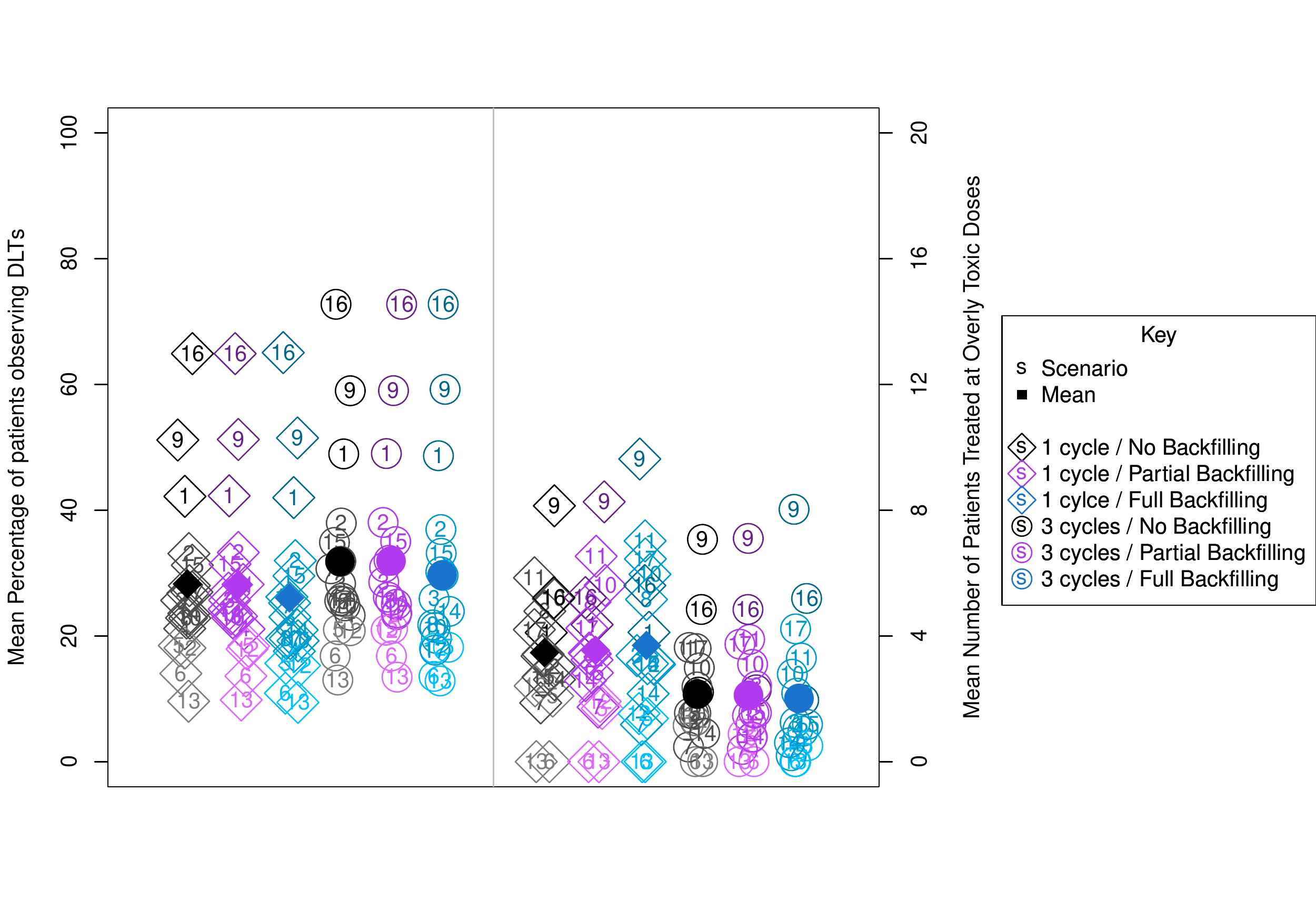}
     \caption{Mean number of DLTs and mean number of patients treated at overly toxic doses across scenarios. Darker colours indicate scenarios where a higher number of doses are unsafe.} \label{fig:DLT_overs}
  \end{figure}

The other important metrics used in dose-finding trials concern the safety of the patients within the trial. Figure~\ref{fig:DLT_overs} shows the distribution of mean percentage of DLT responses and the mean number of patients assigned to overly toxic doses. Although backfilling increases the number of DLT responses observed in every scenario, the percentage of patients observing a DLT response decreases slightly when full backfilling is used, and is largely unchanged for partial backfilling.  For the most part, the increase in numbers of patients exposed to overly toxic doses are in scenarios where all doses are unsafe. Hence one mistaken escalation increases this exposure by three cohorts when backfilling instead of one when not. The benefit in reducing the number of patients exposed to overly toxic doses comes in scenarios where the MTD is in the middle of the dose range, and the backfilling expansion provides more information on the lower doses, resulting in more cautious escalation.\\

Interestingly, the patterns observed in the setting with one cycle of follow-up are not all duplicated in the setting with three cycles of follow-up. It is not the case here that the PCS and PAS increase across all scenarios when backfilling is employed. In fact in some scenarios, there is a noticeable decrease in PCS when backfilling is employed. For example, in scenarios 5 -- 7, where the PCS is low in both settings, it is lower when backfilling is used. This reflects further the point raised earlier that the backfilling leads to a more cautious escalation. In most cases, the PAS is higher when backfilling is implemented.\\

In terms of mean sample size and trial duration, the circles on Figure~\ref{fig:SSize_Durations} display this relationship. This relationship is similar to that observed with one cycle. For every additional patient, the trial duration decreases by 0.3 weeks, for both full and partial backfilling. Again, the magnitude of this overall increase or decrease is larger for fully backfilling. Likewise, the comparisons of the safety aspects of the trials are similar. With three cycles of observation, fully backfilling increases the total number of DLTs in all scenarios, decreases the percentage of DLT responses overall, and decreases the number of patients exposed to overly toxic doses in some scenarios, and increases in others. On average across scenarios, there is a slight decrease in the number of patients exposed to overly toxic doses. \\

\subsubsection{Calibrated Prior}
Now that we have investigated the behaviour of each approach when the same prior specification is used, we move to look at the potential of each approach when the prior is calibrated for the design. Note that for the case of partial backfilling, the calibrated prior with full backfilling is used.\\

\begin{figure}
       \includegraphics[width=1\linewidth]{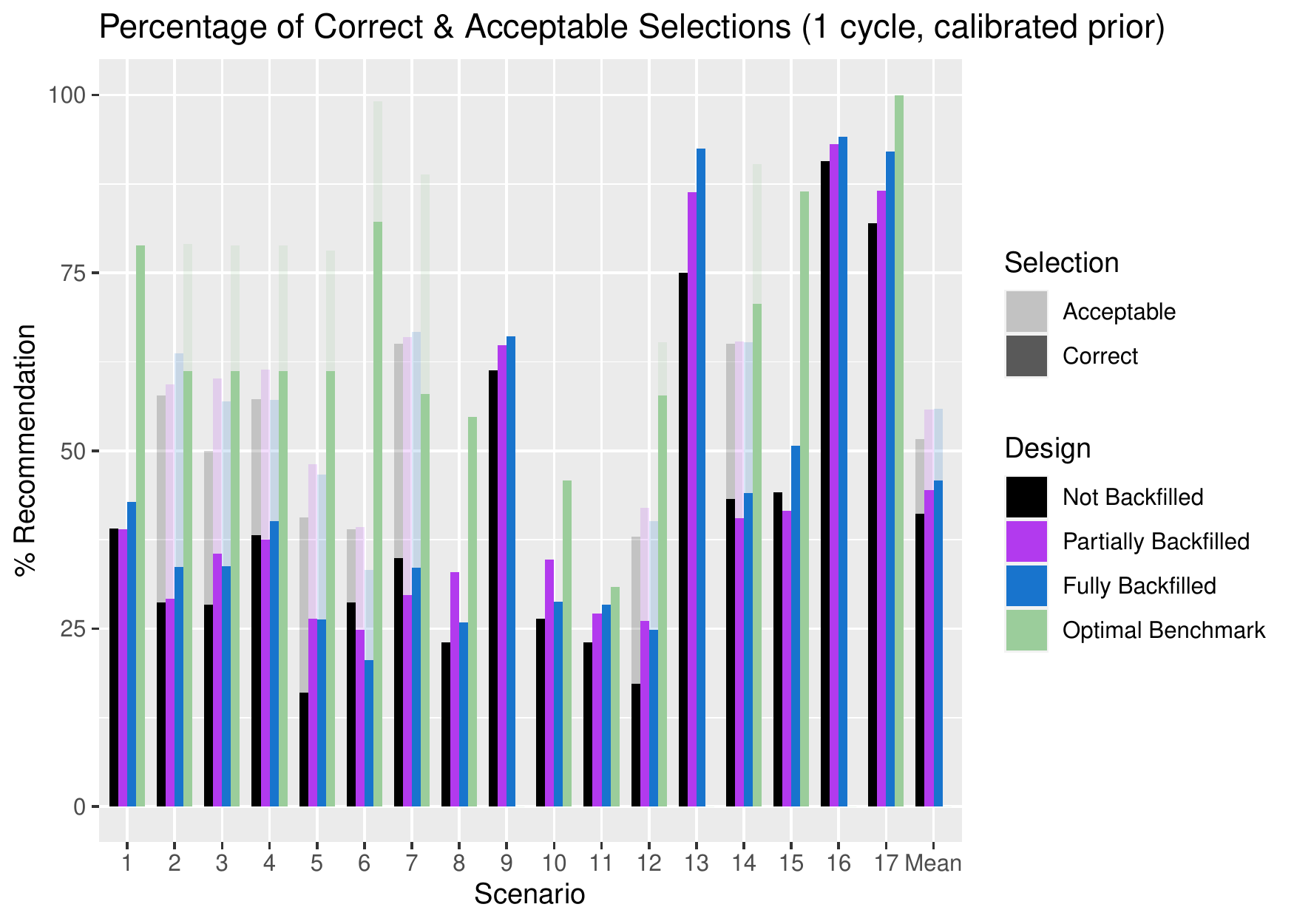}
     \caption{Proportion of correct and acceptable selections across scenarios for one cycle of follow-up, using calibrated priors.}   \label{fig:PCSPAS_cyc1_cal}
  \end{figure}

\begin{figure}
   
      \includegraphics[width=1\linewidth]{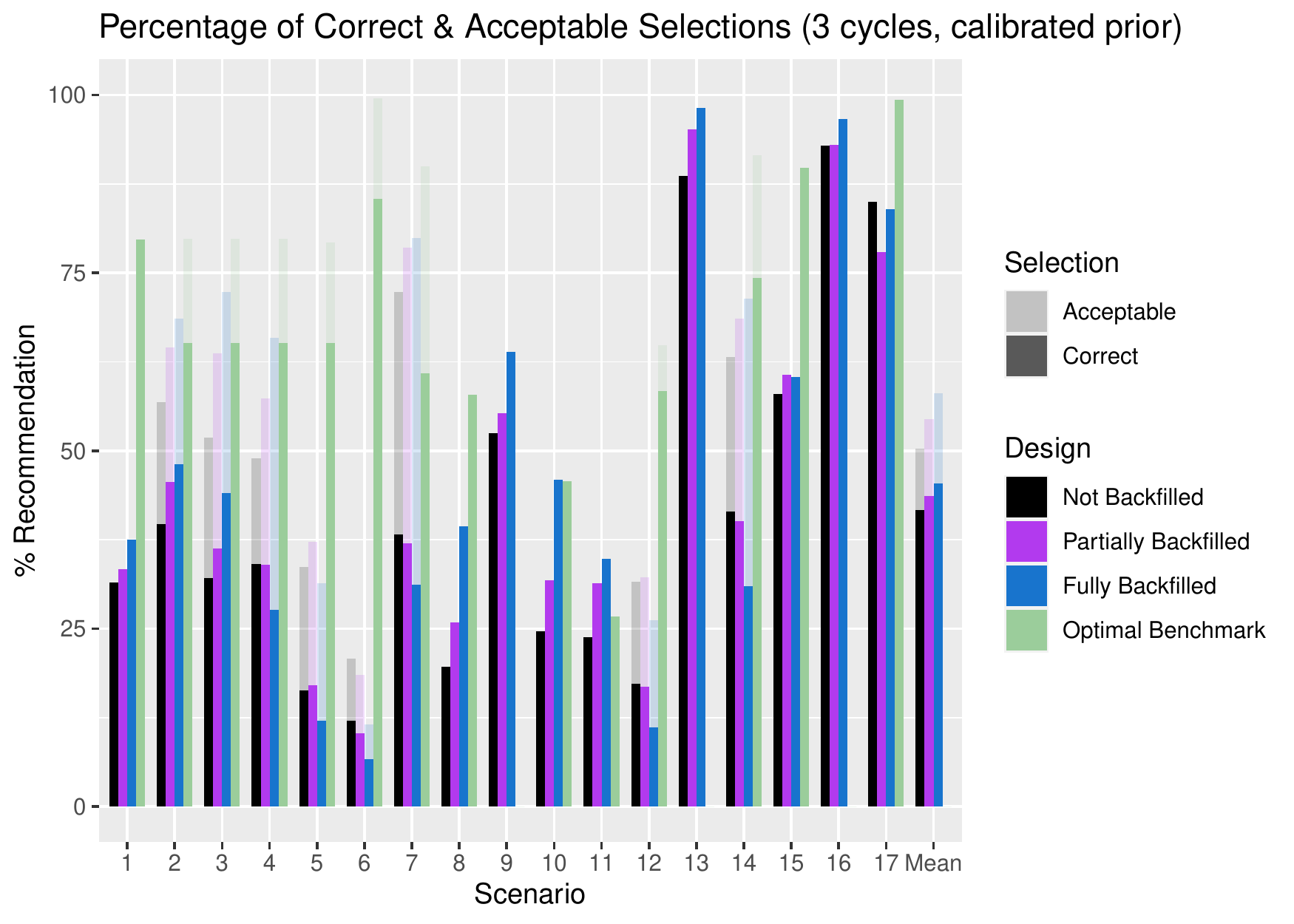}
     \caption{Proportion of correct and acceptable selections across scenarios for three cycles of follow-up, using calibrated priors.} \label{fig:PCSPAS_cyc3_cal}
  \end{figure}

Figures \ref{fig:PCSPAS_cyc1_cal} and \ref{fig:PCSPAS_cyc3_cal} illustrate the proportion of correct and acceptable selections in the settings with a toxicity assessment period of one and three cycles, respectively. All but scenarios 6 and 7 see an increase in the proportion of correct selections when full backfilling is employed in the setting using one cycle. In most scenarios the partial backfilling gives a level of performance in between the full backfilling and no backfilling as before. However, interestingly in some scenarios, such as scenarios 3, 8 and 10, partially backfilling gives a higher proportion of correct and acceptable selections than both when using full and no backfilling. In these scenarios, the MTD is the third dose level. When the MTD is the fourth dose level, partially backfilling performs worse than the other two approaches. This can be explained by the calibration which assumes full backfilling. But since full and partial backfilling result in different levels of information at lower doses, the calibration is no longer optimal for partial backfilling. From this we can deduce that calibrating the prior for the setting at hand has distinct advantages. \\

When the DLT period is three cycles, a larger difference between backfilling and not backfilling can be seen. In most cases, full backfilling increases the proportion of correct selections (up to 21\% in scenario 10). In some cases, however, backfilling again does not increase the proportion of correct selections (e.g. scenarios 5 and 6 where the MTD is in the higher dose range). Partial backfilling once more gives a performance level in between no and full backfilling as previously.\\

\begin{figure}
   
      \includegraphics[width=1\linewidth]{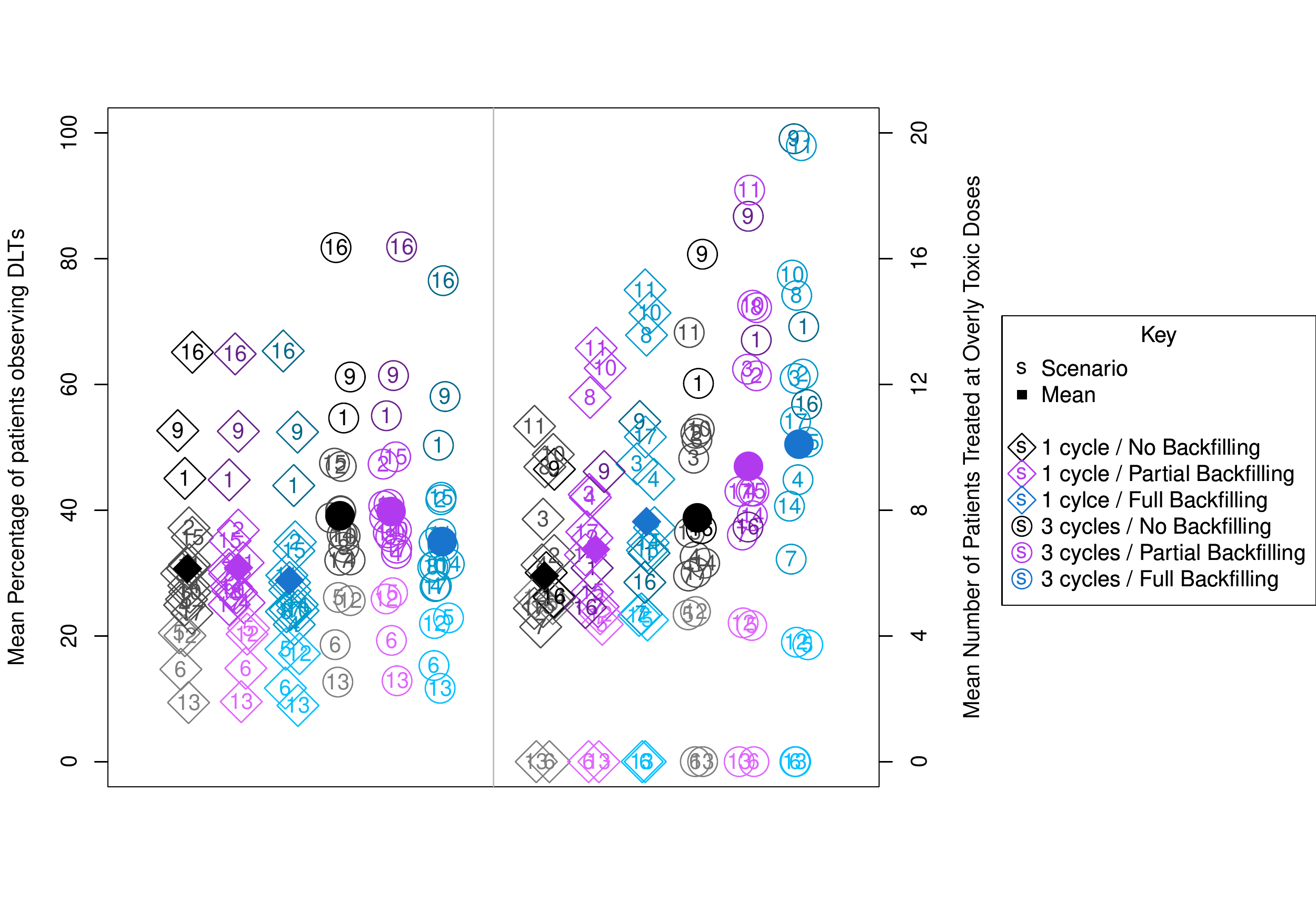}
     \caption{Mean number of DLTs and mean number of patients treated at overly toxic doses across scenarios. Darker colours indicate scenarios where a higher number of doses are unsafe, using calibrated priors.} \label{fig:DLT_overs_cal}
  \end{figure}

Figure \ref{fig:DLT_overs_cal} shows the percentage of patients who experience DLT responses, and the number of patients treated at overly toxic doses. When backfilling is employed, more patients are on average treated at overly toxic doses. The exceptions are scenarios 5 and 12, where the fifth highest dose is the MTD and therefore there is only one overly toxic dose. In terms of percentage of patients observing DLTs, this is less than or equal in every scenario for the use of backfilling. The percentage of patients observing a DLT response is less when backfilling is employed.\\

The mean trial duration is drastically reduced with the use of full backfilling, on average by seven weeks. The largest reduction of 10 weeks is seen in scenarios 4 and 10, when the MTD is in the middle of the dose range, for both a DLT period of one and three cycles. These are also scenarios where the mean total sample size increases by a large amount. The mean total sample size increases on average 16 patients when one cycle of follow-up is used, and 18 patients when three is used. For every additional patient, over the 17 considered scenarios, the trial duration reduces by 0.4 weeks for both one and three cycles of follow-up. \\

Interestingly, the use of partial backfilling on average slightly increases the trial duration over the setting with no backfilling when one cycle of follow-up is used, driven by an increase in duration in scenarios where the MTD is mid dose-range. This is once again a consequence of the prior that is calibrated for full backfilling. \\

     \begin{figure}
  
      \includegraphics[width=1\linewidth]{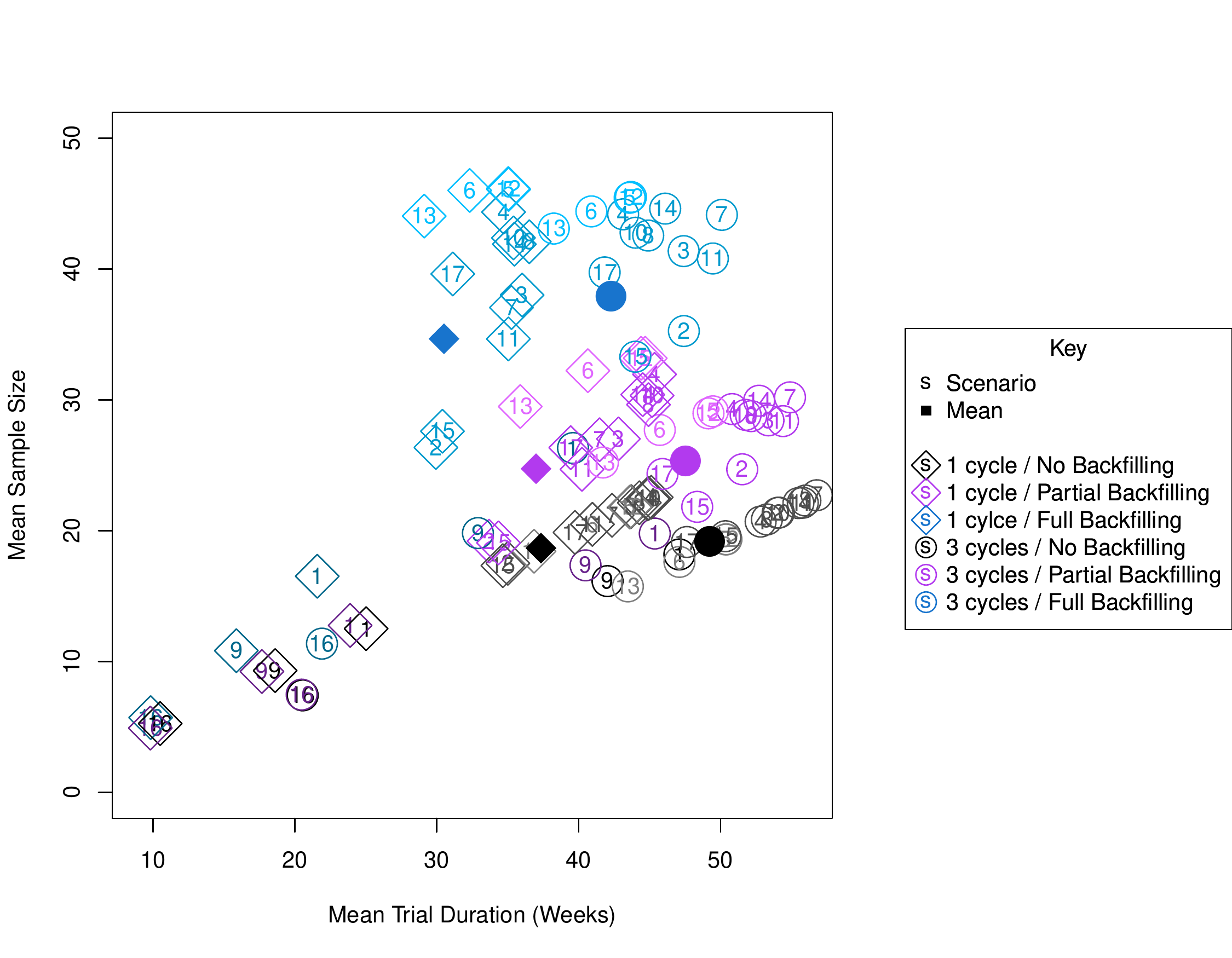}
     \caption{Mean total sample size and mean trial duration across scenarios. Darker colours indicate scenarios where a higher number of doses are unsafe, using calibrated priors.}  \label{fig:SSize_Durations_cal}
  \end{figure}

\section{Discussion} \label{sec:Discussion}

In this work, we have investigated the effect of backfilling on the operating characteristics of dose-escalation studies. The main reasons to utilize backfilling include to gain better understanding of the safety, tolerability and activity of the treatment under investigation and to aid determining the recommended Phase II dose. In this work we have focused on the implications of backfilling on the estimation of the MTD and the duration of the study. We found that backfilling increases the chance of identifying the MTD while reducing the duration of the study. This comes at the cost of an increased number of patients required in studies that use backfilling. The impact of backfilling on the accuracy is larger in the setting with a DLT assessment period of one cycle then when three cycles of follow-up are used. \\

In our evaluations, patients for backfilling are available for recruitment immediately which clearly is an optimistic assumption. Moreover the use of backfilling does increase the number of patients in the study. Nevertheless we did see consistent benefits of backfilling. Specifically, one additional patient did yield a reduction of the study duration of approximately half a week.\\

Our investigations explore two different settings with respect to the prior distributions used. In the first setting the prior is chosen with a study in mind that does not plan to use backfilling, while the second considers backfilling as an option at the outset. Encouragingly we find that, irrespective of the setting, the benefits of backfilling on estimation of the MTD and duration of the study are fairly consistent.

\section*{Data Availability Statement}
All data is simulated according to the specifications described.

\section*{Acknowledgements}
This report is supported by the NIHR Cambridge Biomedical Research Centre (BRC-1215-20014). The views expressed in this publication are those of the authors and not necessarily those of the NHS, the National Institute for Health Research or the Department of Health and Social Care (DHSC). T Jaki and H Barnett received funding from UK Medical Research Council (MC\_UU\_00002/14).

\bibliography{backfill_bib}
\end{document}